\documentclass{aastex631}
\usepackage{lineno}
\usepackage{soulutf8}
\usepackage{ulem}
\usepackage[encoding,filenameencoding=utf8]{grffile}

\def\ltsima{$\; \buildrel < \over \sim \;$}
\def\gtsima{$\; \buildrel > \over \sim \;$}
\def\lsim{\lower.5ex\hbox{\ltsima}}
\def\gsim{\lower.5ex\hbox{\gtsima}}

\shorttitle{New parameters to trace the dynamical evolution of star clusters}
\shortauthors{Bhat et al.}
\graphicspath{{./}{figures/}}

\begin{document}

\title{New parameters for star cluster dynamics: the effect of primordial binaries and dark remnants}

\author{B. Bhat}
\affiliation{Dept. of Physics and Astronomy ``A. Righi'', University
  of Bologna, Via Gobetti 93/2, Bologna, Italy}
\affiliation{INAF Osservatorio di Astrofisica e Scienza dello Spazio
  di Bologna, Via Gobetti 93/3, Bologna, Italy}

\author{B. Lanzoni}
\affiliation{Dept. of Physics and Astronomy ``A. Righi'', University
  of Bologna, Via Gobetti 93/2, Bologna, Italy}
\affiliation{INAF Osservatorio di Astrofisica e Scienza dello Spazio
  di Bologna, Via Gobetti 93/3, Bologna, Italy}

\author{F. R. Ferraro}
\affiliation{Dept. of Physics and Astronomy ``A. Righi'', University
  of Bologna, Via Gobetti 93/2, Bologna, Italy}
\affiliation{INAF Osservatorio di Astrofisica e Scienza dello Spazio
  di Bologna, Via Gobetti 93/3, Bologna, Italy}

\author{E. Vesperini}
\affiliation{Dept. of Astronomy, Indiana University, Bloomington, IN 47401, USA}

\begin{abstract}
By studying the normalized cumulative radial distribution (nCRD) of the stars in the central region of a Monte Carlo-simulated globular cluster, we recently defined three parameters able to pinpoint the stage of internal dynamical evolution reached by the system: $A_5$ (i.e., the area subtended by the nCRD within 5\% the half-mass radius, $r_h$), $P_5$ (the value of the nCRD at 5\% $r_h$), and $S_{2.5}$ (the slope of the nCRD at 2.5\% $r_h$). Here we extend the analysis and explore the effects that different fractions (0\%, 10\%, and 20\%) of primordial binaries and stellar-mass black holes (BHs)  induce on the dynamical history of the system. As expected, the gradual contraction of the cluster becomes milder and core collapse shallower for increasing binary fraction. Nevertheless, the cluster dynamical evolution is still properly traced by the three parameters. For models with a larger initial retention of stellar mass BHs the evolution depends on the timescale of their subsequent dynamical ejection. An early dynamical ejection of BHs results in a long-term evolution of the three parameters similar to that found in systems with no initial BH retention. Conversely, in the model that retains a large number of BHs for extended time (slow dynamical ejection of BHs), the system is characterized by a less concentrated structure and by the lack of significant temporal evolution of the three parameters.  The smaller values of the three parameters found in this case might be used to indirectly infer the possible presence of BHs in the cluster.
\end{abstract}


\section{Introduction}
\label{sec:intro}
Globular Clusters (GCs) are gas-free stellar systems that can harbour
up to a few million stars of different masses. They are the most
populous and oldest systems where stars can be individually
observed. At odds with what happens in galaxies, where the orbital
motion of stars primarily depends on the average gravitational
potential, GCs are ``dynamically active'' (collisional) systems, where
two-body interactions cause kinetic energy exchanges among stars and
gravitational perturbations to their orbits, bringing the cluster
toward a nearly thermodynamically relaxed state in a timescale (the
relaxation time) that can be significantly shorter than their
age. Because of such interactions, heavy stars tend to progressively
sink toward the central region of the cluster (dynamical friction),
while low-mass stars preferentially escape from the system. This
yields a progressive contraction of the core, producing a significant
increase of its density: the so-called ``core-collapse'' (CC).  The
runaway contraction is halted by the energy provided by the binary
systems, and the post-CC phase is characterized by oscillations in the
core size and central density.  The recurrent gravitational
interactions among stars thus progressively modify the structure of
the system at a rate that depends in a very complex way on the initial
and the local conditions (such as the total cluster mass, the central
density, the fraction of binaries and dark remnants, as well as the
orbit within the host galaxy). Hence, in spite of a comparable
chronological age, different clusters can have very different
dynamical ages \citep[e.g.,][]{meylan+97}.
Determining the dynamical evolutionary stage of star clusters is
  important not only to have a complete physical understanding of
  these systems, but also because internal dynamic processes can have
  a significant impact on their stellar population and observational
  properties. For instance, blue straggler stars and millisecond
  pulsars are not predicted by the stellar evolution models of single objects, but they
  are originated by dynamical processes involving direct stellar
  collisions and/or the evolution of binary systems
  \citep[e.g.,][]{mccrea64, hills_day76, bhattacharya+91,
    sills+97}. Hence, their frequency and their properties depend on
  the dynamical stage of the system and can be used to get information
  on the internal dynamics of GCs \cite[see, e.g.,][]{freire+04,
    ransom+05, ferraro+09, verbunt_freire14, ferraro+18, ferraro+19}.
 In addition, other events such as neutron star and white dwarf mergers are thought to be enhanced in core-collapsed clusters, with important implications for our understanding not only of the internal dynamics of the system itself, but also, for example, of the rate of type Ia supernovae, and the origin of short gamma-ray bursts (\citealp[see, e.g.,][see discussion in ]{Grindlay+06,rodriguez+16,kremer+21}\citealt{ye+20}). 
The approach most commonly adopted to estimate the
clusters' dynamical ages and establish their current dynamical phase
relies on the properties of the projected density profile of the
system and its comparison with \citet{king66} models. The study of the
cluster density profile may reveal features associated with, for
example, the CC or the post-CC phase and allows the determination of
relaxation timescale for the cluster inner regions (e.g.,
\citealt{djorgovski+84, djorgovski93, ferraro+03, ferraro+09}). These
diagnostics, however, have been found to be not fully reliable and
univocal in properly assessing the level of dynamical evolution
reached by star clusters, and alternative indicators have been
proposed in the recent years, either based on peculiar populations of
heavy stars (as blue stragglers) that are sensitive tracers of the
dynamical friction efficiency \citep[e.g.,][]{ferraro+12, ferraro+18,
  ferraro+19, lanzoni+16}, or based on the internal mass and kinematic
structure of each system \citep[e.g.,][] {baumgardt_makino03,
  tiongco+2016, bianchini+2016, webb_vesperini2017, bianchini+2018,
  bhat+22}. In particular, in \citet[][hereafter, B22]{bhat+22} we
used a first set of Monte Carlo simulations following the evolution of
a GC well beyond CC, to define three new empirical parameters able to
measure the dynamical age of the system from the normalized cumulative
radial distribution (nCRD) of the stellar population in the central
region of the cluster.  Specifically, we observed that the shape of
the nCRD progressively varies with time according to the system's
dynamical evolution, and we defined three parameters that appear to
efficiently trace and quantify these morphological changes: they have
been named $A_5$, $P_5$, and $S_{2.5}$ (see Section \ref{sec:A5etc}).
The investigation presented in B22 shows that these nCRD-based
parameters progressively increase with time, reach their maximum
values at CC, and then exhibit some oscillations during the post-CC
phase, closely tracing the dynamical evolution of the system.  Hence,
they appear to be very promising empirical tools for assessing the
dynamical evolutionary stage reached by star clusters.  However, that
first study explored a limited set of initial conditions not including
primordial binaries or a significant fraction of dark remnants, as
black holes and/or neutron stars. These objects are observed in GCs
 \citep[e.g.,][and references therein]{milone+12, sollima+12,
    Strader+12, Chomiuk+13, giesers+18, giesers+19} and are known to
play a relevant role in the dynamics of stellar systems.  In fact, the
``heating effect'' of these objects weakens (and then halts) the core
contraction \citep[see, e.g.,][]{vesperini+94, trenti+07,
  chatterjee+10}, and they are also thought to substantially affect
the structural properties of the host system
\citep[e.g.,][]{sigurdsson+93, mackey+07, mackey+08, morscher+15},
also quenching mass segregation \citep[see, e.g.,][]{Gill+08,
  alessandrini+16}.

The present work is thus specifically aimed at quantifying the effect
of primordial binaries and stellar-mass black holes on the time
evolution of the three parameters defined in B22. The paper is
organized as follows. In Section \ref{sec:simu} we describe the
initial conditions of the Monte Carlo simulation runs and the
methodology adopted in the following analysis. In Section
\ref{sec:dens} we present the determination of the projected star
density profile and the adopted King fit procedure. Section
\ref{sec:newdyn} describes the method and assumptions used for the
construction of the nCRDs, and the definition and properties of the
three nCRD dynamical indicators in the case of three different
fractions of primordial binaries.  Section \ref{sec:dark_remnants} is
devoted to the analysis of the effects induced by a sub-system of
black holes.  The summary and conclusions of the work are discussed in
Section \ref{sec:discussion}.

\section{Methods and Initial conditions}
\label{sec:simu}
In this paper, we use five Monte Carlo simulations performed with
MOCCA code \citep{Hypki_Giersz_2013,Mocca_giersz} to thoroughly follow
the dynamical evolution of GCs with different primordial binary
fractions and dark remnant retention fractions. In addition to
two-body relaxation and tidal truncation effects, the code also models
stellar and binary evolution by means of the SSE and BSE codes
\citep{Hurley_2000, Hurley_2002}, thus providing, for each star at any
evolutionary time, not only the position and the velocity, but also
the mass, and the $B-$ and $V-$ band magnitudes.  One of the
simulation is the same as presented in B22, and it is used here
for comparison purposes.  It initially has 500K stars with masses
ranging between $0.1 M_\odot$ and $100 M_\odot$, following a
\citet{IMF} mass function, and distributed as a \citet{king66} model
with dimensionless central potential $W_0 = 6$. Supernova kicks are
assumed to follow a Maxwellian distribution with dispersion equal to
265 km s$^{-1}$ \citep{Hobbs_05}. The cluster is tidally underfilling,
with a three-dimensional half-mass radius $r_h = 2$ pc and a Jacobi
radius equal to 61 pc (corresponding to the value that the system
would initially have if orbiting at a Galactocentric distance $R_g=$ 
4 kpc). No primordial binaries are included in this run, which will be
referred as ``BF0'' throughout the paper.  Two other simulations have
been performed using the same initial conditions except for their
primordial binary fraction;  these simulations start with a total
  number of single and binary stars, $N=N_s+N_b=500$ K and binary
  fraction, $N_b/(N_s+N_b)$ equal to 10\% and 20\% in the runs that
  will be referred to as ``BF10'' and ``BF20'', respectively. The
  distribution of binary properties (as mass ratio, period and
  eccentricity) are set according to the eigenevolution procedure
  described in \citet{kroupa95} and \citet{kroupa+2013}.  The same
initial conditions as those of the BF0 model, but with a reduced kick
velocity for stellar black holes, have been used also in the fourth
simulation (which we name ``DRr''). The fifth simulation, hereafter referred to as ``DRe'', has slightly different initial conditions: the initial cluster density profile follows that of a King model with $W_0=$ 7,  $r_h=$ 1 pc, $R_g=$ 2 kpc. It starts with 10\% primordial binary fraction and similar to the DRr run, it has a reduced kick velocity for the stellar mass black holes. With the DRr and DRe simulations we explore the effect of dark remnants (retained black holes) on the cluster evolution and the time dependence of the three parameters introduced in B22. The initial conditions of all the simulations are listed in the Table \ref{tab:list of sims}.	

\begin{figure}
\centering
\includegraphics[scale=0.4]{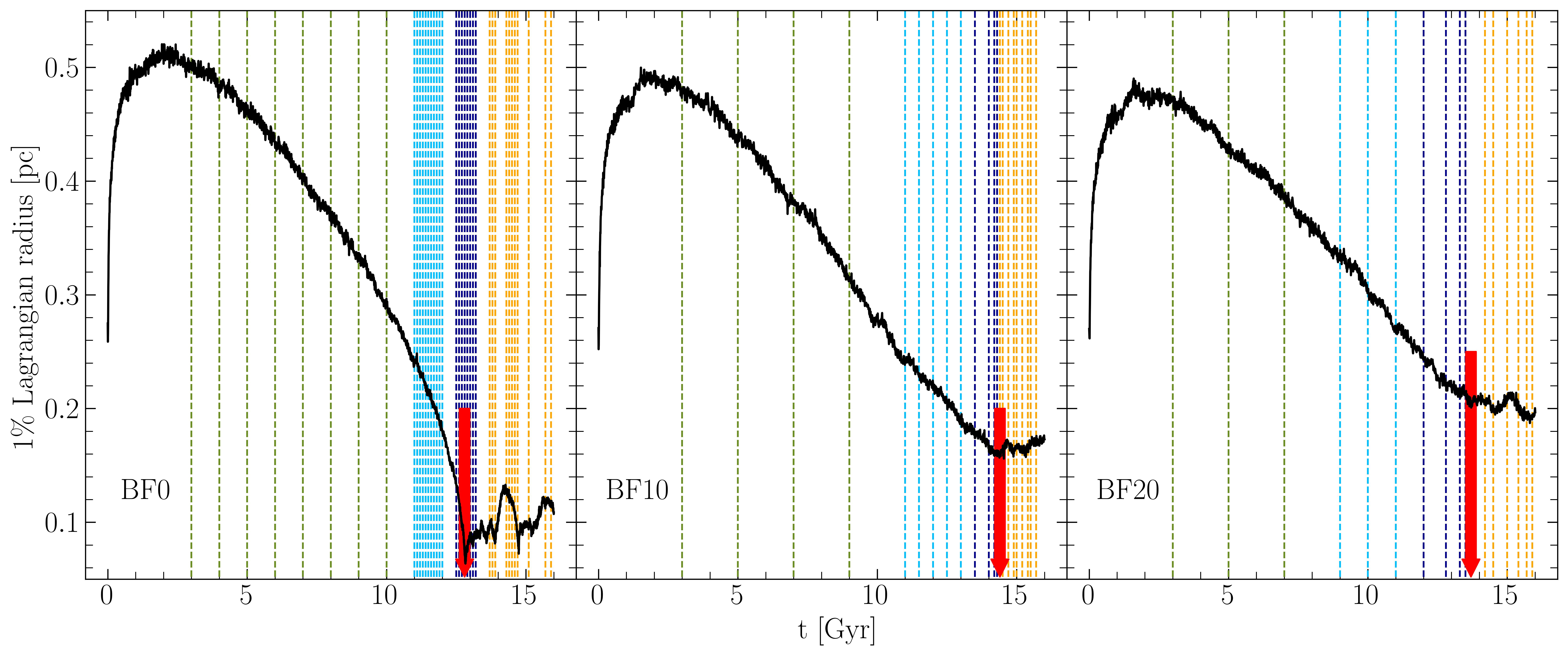}
\caption{Time evolution of the 1\% Lagrangian radius (black line) of
  the three simulations differing only in their primordial binary
  fraction: no primordial binaries in run BF0 (left panel), 10\% and
  20\% initial binary fractions in simulations BF10 and BF20 in the
  central and right panels, respectively. The vertical lines in each
  panel mark the time snapshots extracted and analyzed in each run,
  color-coded following the different evolutionary stages of the
  system: green, cyan, blue and yellow colors for early, pre-CC, CC,
  and post-CC phases, respectively. The time of CC ($t_{\rm CC}=12.8,
  14.4$, and 13.7 Gyr for BF0, BF10, and BF20, respectively)
  is marked with a large red arrow in all the panels.}
\label{fig:lagrange_rad}
\end{figure}

While the analysis of the DRr and DRe simulations is addressed separately in
Section \ref{sec:dark_remnants}, Figure \ref{fig:lagrange_rad} shows
the time evolution of the 1\% Lagrangian radius ($r_{1\%}$, i.e., the
radius including 1\% of the total cluster mass) for the BF0, BF10, and
BF20 runs.  This clearly illustrates the evolution of the cluster's
inner structure and how it is affected by the presence of binary
systems.  In all cases, an initial expansion (driven by heavy mass
loss from young, massive stars) is followed by a phase where two-body
relaxation becomes dominant. This leads to a progressive contraction
of the core, which culminates in the CC event when $r_{1\%}$ reaches
the lowest value (red arrows in the figure). The CC time, $t_{CC}$,  is approximately equal to $t_{\rm CC}=12.8$ Gyr, $t_{\rm CC}=14.4$ Gyr, and $t_{\rm CC}=$ 13.7 Gyr for
the BF0, BF10, and BF20 simulations, respectively.  As apparent and as
expected, the overall effect of primordial binaries is to reduce the
depth of CC and quench the post-CC gravothermal oscillations. 
In fact, if no primordial binaries are present, the system undergoes a phase of deep CC, until enough binaries are dynamically formed (see \citealp{Mocca_giersz} and \citealp{giersz+01} for details on the procedure for dynamical binary formation in MOCCA) in the core and stop the contraction.  Conversely, in a cluster with
substantial primordial binaries, the core contraction is hindered 
  by binary burning (binaries acting as energy sources), which
prevents the system from undergoing deep CC.  Indeed, Figure
\ref{fig:lagrange_rad} shows that the rapid phase of deep contraction
immediately before CC is almost bypassed, and the CC event is reached
in a smoother way in the BF10 and BF20 runs. While $r_{1\%}$ shrinks
by a factor of 8 in the BF0 simulation, the factor is reduced to $\sim
3$ in the BF10 run, and to 2.5 in the case of 20\% primordial
binaries.  A clear phase of gravothermal oscillations, during which
$r_{1\%}$ undergoes cyclic expansions and contractions, is well
distinguishable after CC in the BF0 simulation, while it is absent in
our simulations with primordial binaries.

To more quantitatively examine the impact of primordial binaries on
the cluster dynamical evolution, in the following sections we extend
the same analysis presented in B22 (for the BF0 run) to the BF10 and
BF20 cases.  Here we summarize the main aspects of the work, while
more details about the adopted procedures can be found in the previous
paper.  As in B22, we have extracted different time snapshots
corresponding to various phases of the cluster dynamical evolution in
the three runs.  They are marked in Figure \ref{fig:lagrange_rad} with
vertical dashed lines, color-coded as follows: {\it green} color for
the early slow contraction phase, {\it cyan} for the subsequent phase
leading to CC, {\it blue} for the CC phase, and {\it yellow} for the
the post-CC phase.  Following the approach adopted in B22, every
snapshot is assumed to be a possible configuration of a real cluster
observed in different stages of its dynamical evolution, and to be as
much as possible consistent with real cases, the analysis of the
simulated data has been done from the point of view of an
observer. Thus, standard procedures and approximations adopted in
observational works have been applied: in each snapshot, the simulated
cluster is projected onto a 2D plane, and a distance of 10 kpc from
the observer has been assumed to transform the distances from the
centre of the system from parsecs to arcseconds. In addition,
  binary systems have been treated as ``stellar blends'', consistently
  with the fact that the two components cannot be individually
  resolved at the distances of Galactic GCs. Hence, the magnitude of each binary system has been determined by summing up the
  luminosities of the two stellar components.

\begin{table}
\caption{\label{tab:list of sims}Initial conditions of the simulations.}
\begin{center}
\begin{tabular}{ c c c c c c}
 \hline
 Name & $W_0$ & $\rm{BF}$ & $r_h$(pc) & $R_g$(kpc) & other \\ 
 \hline\hline \\
 BF0  & 6 & 0 & 2 & 4 & ..\\  
 BF10 & 6 & 10 & 2 & 4 & ..\\ 
 BF20 & 6 & 20 & 2 & 4	& ..\\
 DRr  & 6 & 0 & 2 & 4	& Reduced kick velocity for black holes\\
 DRe  & 7 & 10 & 1 & 2	& Reduced kick velocity for black holes\\
 \hline 
\end{tabular}
\end{center}
\tablecomments{Values of the dimensionless central potential ($W_0$), binary fraction (BF), half-mass radius ($r_h$), and Galactocentric distance ($R_g$) adopted as initial conditions in the five Monte Carlo simulations analyzed in this paper (see their name in the first column).}
\end{table}

\section{Projected density profiles}
\label{sec:dens}
\begin{figure}
\centering \includegraphics[scale=0.4,trim={4cm 0 0 0}]{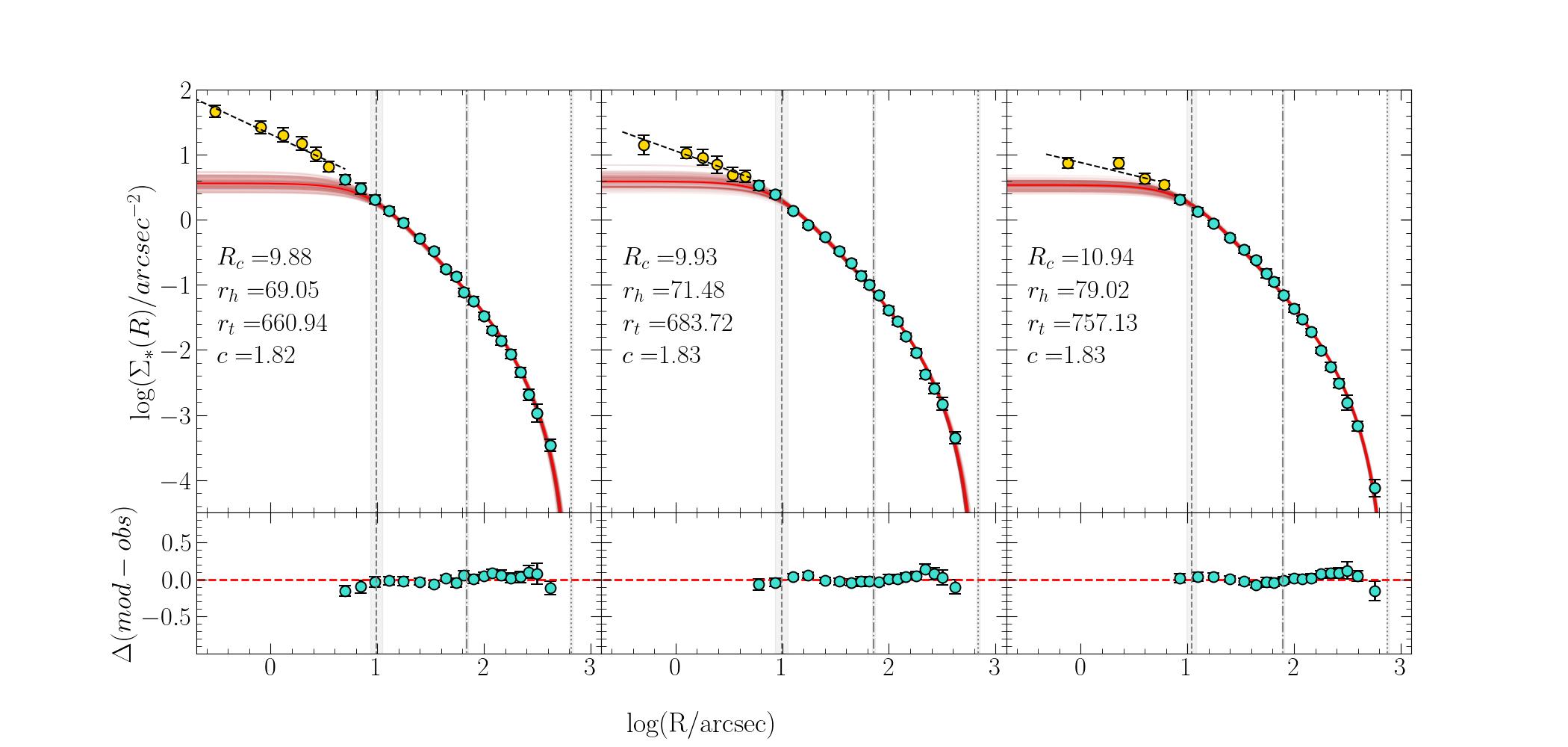}
\caption{Projected density profile for the snapshots extracted at
  $t_{\rm CC}=12.8, 14.4$ and 13.7 Gyr in the BF0 (left panel), BF10
  (central), and BF20 (right) simulations, respectively.  The red
  solid lines correspond to the King models that best-fit the density
  profiles beyond $5\arcsec$ from the center (cyan circles), while the
  innermost cusps (yellow circles) are best-fit by the dashed straight
  lines.  The red shaded regions correspond to the uncertainty of the
  best-fit model. The vertical dashed, dotted-dashed and dotted
    lines mark, respectively, the core, half-mass and truncation radii calculated from the best-fit King models (see also the labels).  The bottom panels show the residual
  between the best-fit model and the ``observed'' density profile.}
\label{fig:king_fit}
\end{figure}

As part of the analysis, we first investigate the effect of different
initial binary fractions on the development and evolution of an inner
density cusp through various epochs of the cluster dynamical
evolution.  To construct the density profile of each extracted
snapshot, we followed the same procedure described in B22 and adopted
in several observational works \citep[e.g.,][]{miocchi+13}. It
essentially consists of counting the number of stars in concentric
annuli around the cluster center, and dividing it by the area of each
radial bin.  We then used a $\chi^2$ method to determine the
\citet{king66} model best-fitting the ``observed'' density profile, by
exploring a grid of models with dimensionless parameter $W_0$ (which
is proportional to the gravitational potential at the center of the
system) varying between 4 and 10.75 in steps of 0.05. This corresponds
to a concentration parameter $c$ spanning the interval between 0.84
and 2.5, with $c$ being defined as the logarithm of the ratio between
the truncation or tidal radius of the system ($r_t$), and the King
radius $r_0$, which is the characteristic scale-length of the model:
$c = log(r_t/r_0)$.  The King model profile well describes the density
distribution of the simulated clusters, except for the snapshots close
to and beyond CC, when a density cusp develops in the center. In these
cases, the model still provides a very good fit to the external
portion of the profile (cyan circles and red lines in Figure
\ref{fig:king_fit}), while the trend in the innermost $\sim 5\arcsec$
is better fitted with a power-law (yellow circles and dashed lines
in the figure).  A density cusp always develops around the CC phase in
all the simulations, and, once formed, it never disappears.  However,
while it is prominent in the BF0 run, it becomes shallower and
sometimes hardly distinguishable in the cases of 10\% and, ever more
so, 20\% primordial binaries.  Thus, the slope of the density cusp
developed in the CC stage is affected by the binary content of the
cluster, consistently with the effect discussed above on the time
evolution of the 1\% Lagrangian radius (Fig. \ref{fig:lagrange_rad}):
star clusters with larger primordial binary fractions experience
shallower CC and develop less steep cusps in the star density profile
at the CC epoch \citep[see also][]{Vesperini+10}.

Of course, while properly reproducing the external portion of the
density profile, the King parameters obtained from the fit after the
exclusion of the innermost $5\arcsec$ cannot be used as an appropriate
description of the overall cluster structure.  In addition, the choice
of $5\arcsec$ is somehow arbitrary, and by changing this value, also
the resulting best-fit King model may change. To overcome these
issues, we thus determined the King models that best-fit the entire
density profile of each snapshot, and we adopted the corresponding
structural parameters in the following analysis (for more details, see
Section 3.1 and Figure 2 in B22).

\section{The nCRD dynamical indicators}
\label{sec:newdyn}
Here we extend the analysis presented in B22 to the case of GCs including a population of primordial binary systems, to test how the newly
defined dynamical indicators, quantifying the morphology of the nCRD
of cluster stars, depend on the primordial binary fraction.  In the
following, we therefore adopt the same methodology fully described in
B22 and aimed at making the three parameters well measurable from
observations.

\subsection{Normalised cumulative radial distribution (nCRD)}
\label{sec:crd}
For each extracted snapshot, to build the nCRD we selected all the
stars brighter than $V_{\rm cut} = V_{\rm TO}+0.5$ (with $V_{\rm TO}$
being the $V-$band magnitude of the main-sequence turn-off point), and
located within a projected distance equal to $0.5 \times r_h$ from the
centre. In the case of binary systems, we considered the combined
  magnitude of the two components, since all binaries remain
  unresolved (they are observed as stellar blends) at the distance of
  GCs. These choices are motivated by the fact that the same
procedure will be applied in future investigations to observational
data, for which proper magnitude selections are needed to avoid
problems of photometric incompleteness, and a common radial cut in
units of a physical scale-length (as $r_h$) is required to allow the
comparison among stellar systems of different intrinsic sizes.  More
specifically, we defined $x=R/r_h$ and considered all the
magnitude-selected stars located between $x=0$ and $x=0.5$. For any value of x, the nCRD is equal to the number of stars within x and is normalized by the total number of selected stars. Hence, by construction,
the nCRD is a curve that monotonically increases from 0 at the centre
($x=0$), to 1 at $x=0.5$. The steeper it is, the more concentrated are
the selected stars toward the center of the system.

\begin{figure}
\centering \includegraphics[scale=0.4]{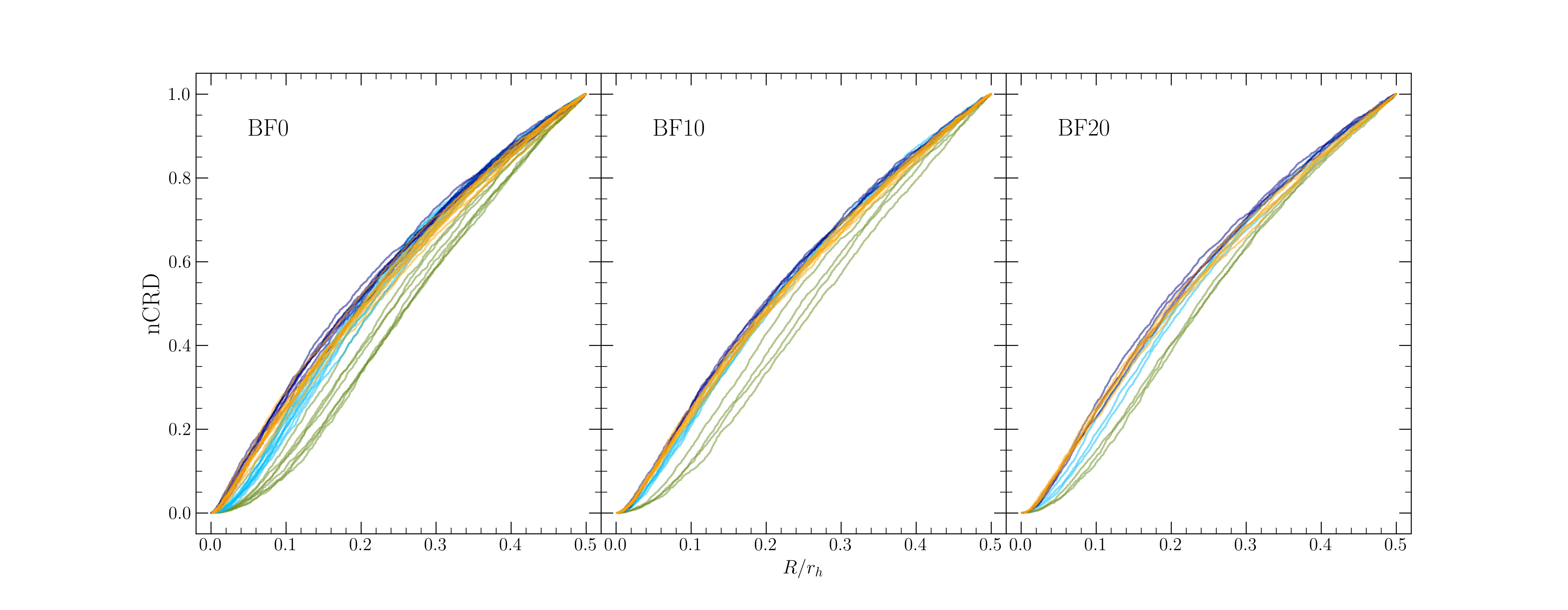}
\caption{Normalised cumulative radial distributions (nCRDs) of stars
  brighter than $V=V_{\rm TO}+0.5$ and located within $0.5 \times r_h$
  from the center, for all the snapshots analysed in the BF0, BF10 and
  BF20 simulations (left, central and right panels, respectively).
  The color coding is the same adopted in Fig.\ref{fig:lagrange_rad}:
  early times (green), pre-CC stages (cyan), CC phase (blue), and
  post-CC epoch (yellow).}
\label{fig:crds}
\end{figure}

Figure \ref{fig:crds} shows the nCRDs thus obtained for all the
snapshots extracted from the three simulations, color-coded as in
Fig. \ref{fig:lagrange_rad}.  Clearly, the morphology of the nCRDs
changes with time following the cluster dynamical evolution, with
shallower curves (less centrally segregated stars) for early
evolutionary times (green lines), and increasingly steeper functions
for more advanced dynamical stages.  These morphological differences
are stronger in the BF0 case, and become progressively less pronounced
in the BF10 and BF20 runs.  This is another manifestation of the
different depth of CC in the three cases (see Figure
\ref{fig:lagrange_rad}). In fact, in the absence of a primordial
population, the binaries present in the cluster are limited to those
dynamically generated in the core during the most advanced stages of
evolution. Hence, they have little effect on the progressive
segregation of cluster stars toward the center.  Conversely, in a
system with a substantial fraction of primordial binaries, binary
burning provides the energy needed to halt core collapse earlier. Thus, the primordial binary fraction does affect the
extent of morphological differences imprinted in the nCRD by the
internal dynamical evolution of the system.

\subsection{The $A_5$, $P_5$ and $S_{2.5}$ parameters}
\label{sec:A5etc}
In B22 we defined\footnote{While the definitions of the three
parameters are exactly the same as in B22, we emphasize that here they
are computed by using distances normalized to the half-mass radius
$x=R/r_h$, while in B22 the adopted normalization was $0.5\times
rh$. This has no impact on the results and on the time evolution of
the parameters; the only difference is that the values of $A_5$
plotted here are half of those published in B22, and those of
$S_{2.5}$ are approximately twice the previous ones.} the following
three parameters quantitatively describing the temporal variations of
the nCRD caused by the cluster dynamical evolution:
\begin{itemize}
\item $A_5$ is the area subtended by each nCRD between the center and
  a projected distance equal to 5\% $r_h$ (hence, between $x=0$ and
  $x=0.05$, with $x=R/r_h$);
\item $P_5$, is the value of the nCRD at 5\% $r_h$ ($x=0.05$), which
  corresponds to the fraction of selected stars located within this
  distance from the centre;
\item $S_{2.5}$ is the slope of the straight line tangent to the nCRD
  at 2.5\% $r_h$ (at $x=0.025$). More specifically, it is the slope of
  the tanget to the third-order polynomial function that best-fits the
  nCRD (the fit being introduced to smooth out the noisy behavior of
  the nCRD itself).
\end{itemize}

They are all defined in the very central region of the cluster, to
best sample the radial distance where the dynamical effects
responsible for the central density growth during the core contraction
and the CC phase are most relevant, and the nCRD morphological
differences are maximized.  The major source of error for the
parameters is the uncertainty in the value of the half-mass radius
derived from the King models. However, as discussed in B22, the size
of these errors is negligible and they are therefore ignored here.

\begin{figure}
\centering \includegraphics[scale=0.4]{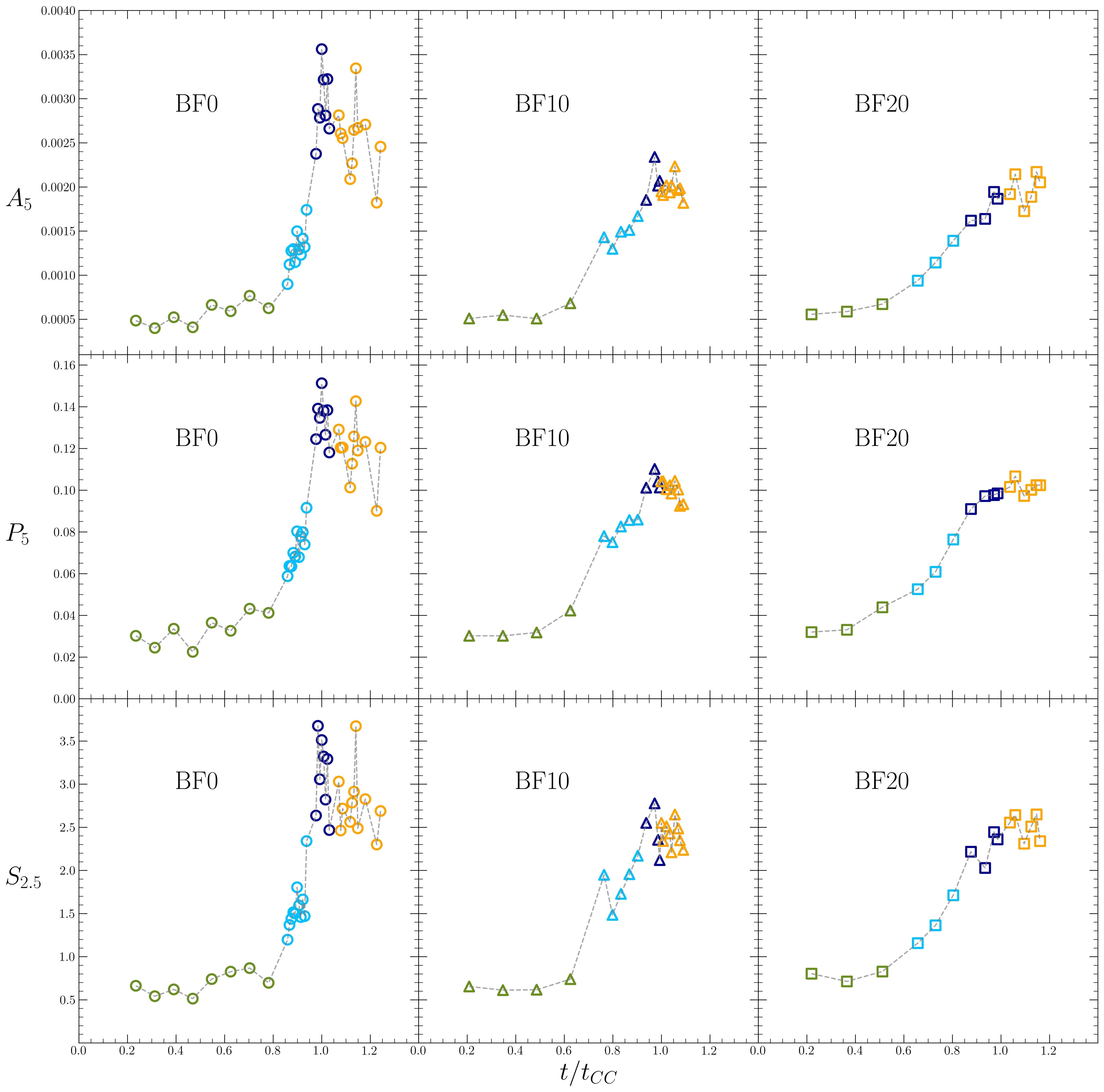}
\caption{ Time evolution of the $A_5$, $P_5$ and $S_{2.5}$
    parameters (top, middle, and bottom panels, respectively) in the
    BF0, BF10, and BF20 simulations (left, central, and right panels,
    respectively). Time is normalised to the CC time of the
  corresponding simulation. The color code is the same as in
  Fig. \ref{fig:lagrange_rad}.}
\label{fig:all_params}
\end{figure}

By construction, these parameters quantifies the evolution of the nCRD
and they increase as a function of the cluster dynamical age. 
  This is clearly shown in Figure \ref{fig:all_params} (top, middle
  and bottom panels for the $A_5$, $P_5$, and $S_{2.5}$ parameters,
  respectively).  In the early dynamical phases (green symbols), each
parameter shows a nearly constant behavior, taking essentially the
same (small) values regardless of the primordial binary
fraction. Then, they increase in the pre-CC phase (cyan symbols),
reach a peak at the CC epoch (blue symbols), and shows some
fluctuations during the post-CC stage (yellow symbols), never receding
to the initial low values.  The gradual growth with time of the
parameters is most pronounced in the BF0 simulation (left panels), and
becomes milder for larger binary fractions (central and right panels).
This is indeed expected for the same reasons discussed above,
reflecting the shallower CC and the lack of gravothermal oscillations
in the models with 10\% or 20\% primordial binary
fraction. Nevertheless, the clear increasing trend with time confirms
the effectiveness of these parameters as proper tracers of the
dynamical aging of the system.

\begin{figure}
\centering \includegraphics[scale=0.4]{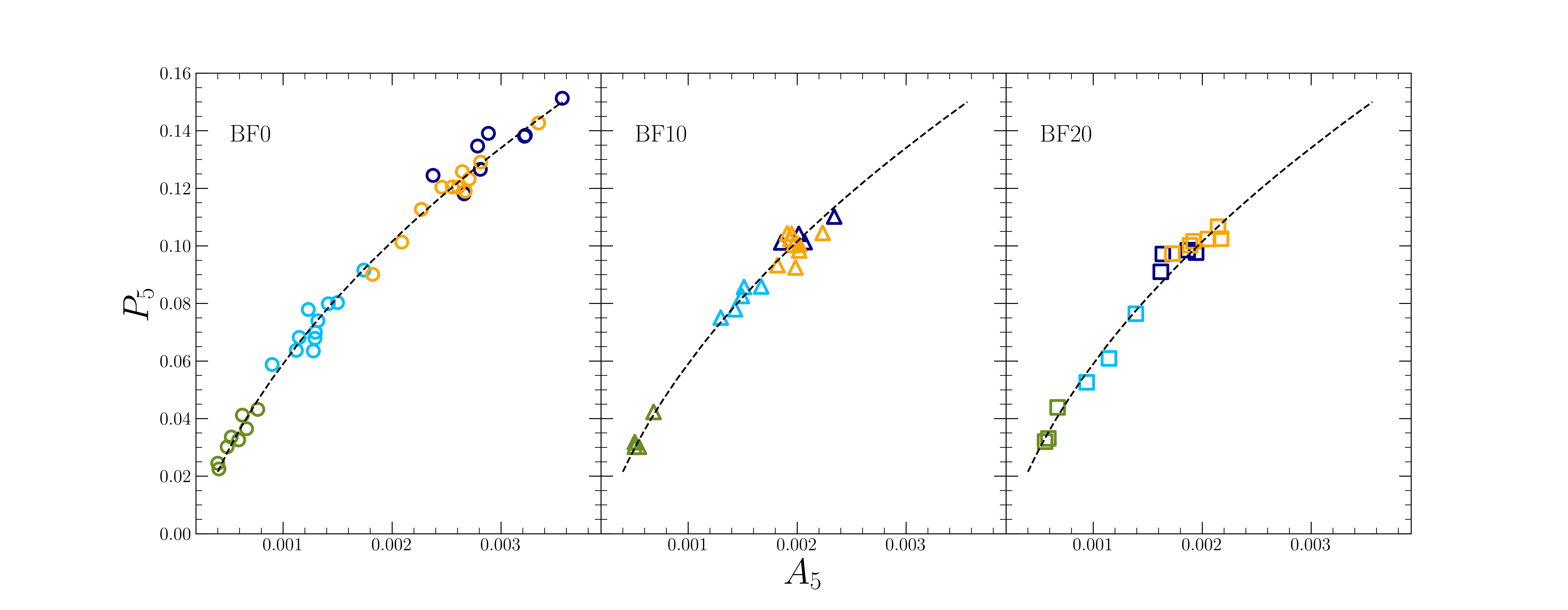}
\caption{$P_5$ parameter plotted against $A_5$ for the BF0, BF10, and
  BF20 simulations (left, central, and right panels,
  respectively). The color code is the same as in all previous
  figures. The black dashed lines are the polynomial fit to the
  distribution obtained in the BF0 run, reported for reference also in
  the central and right panels.}
\label{fig:a5p5}
\end{figure}

\begin{figure}
\centering \includegraphics[scale=0.4]{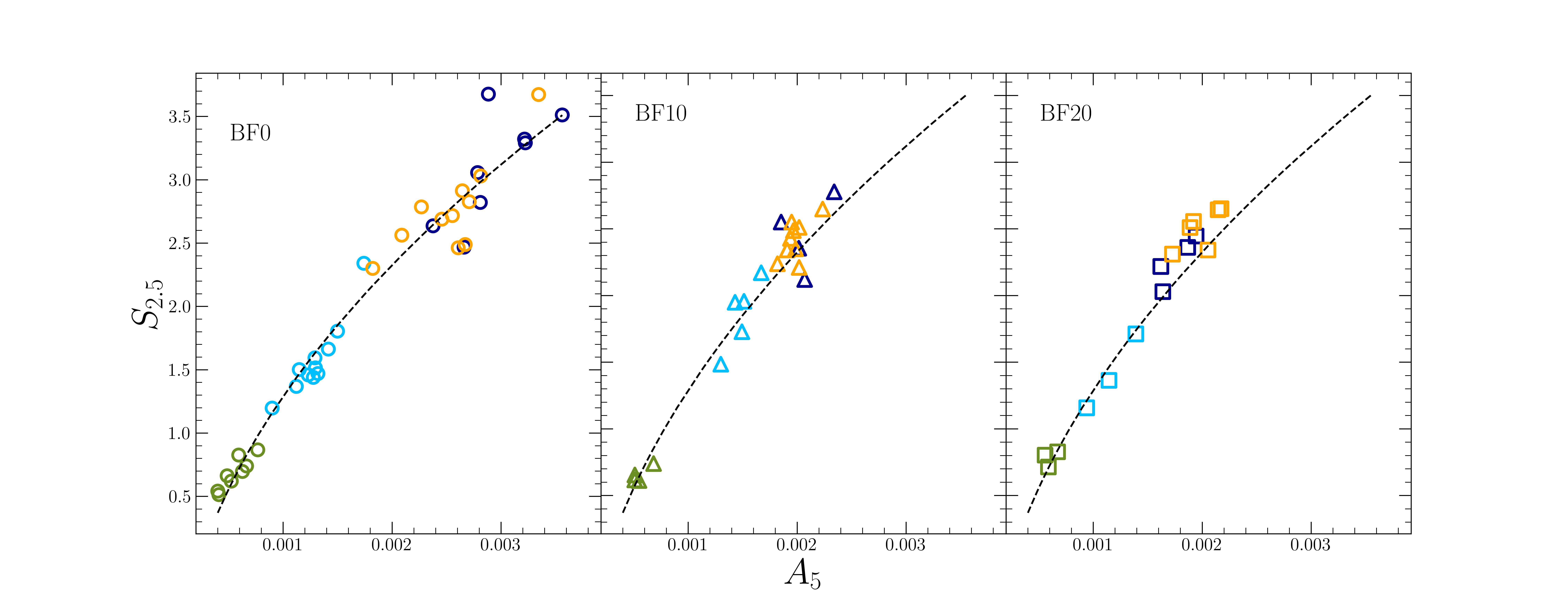}
\caption{As in Fig. \ref{fig:a5p5}, but for the $S_{2.5}$ parameter
  plotted against $A_5$.}
\label{fig:a5s2.5}

\end{figure}

\begin{figure}
\centering \includegraphics[scale=0.4]{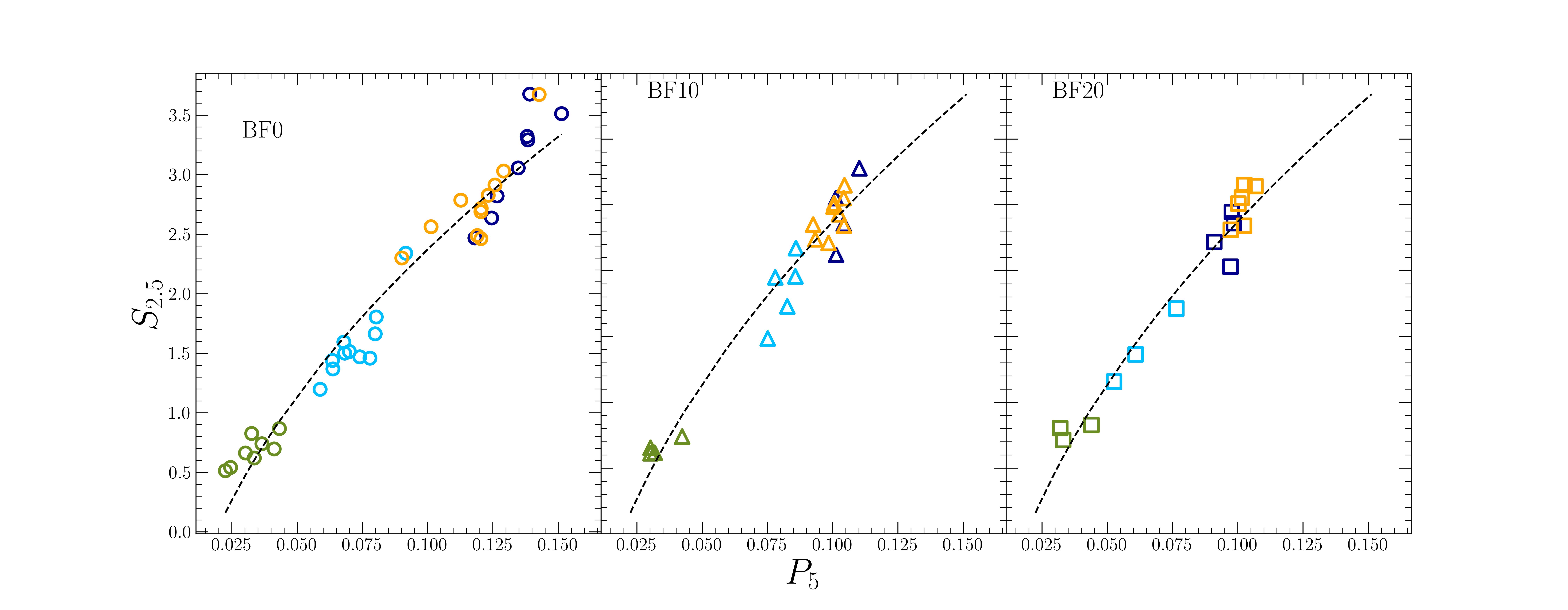}
\caption{As in Fig. \ref{fig:a5p5}, but for the $S_{2.5}$ parameter
  plotted against $P_5$.}
\label{fig:p5s2.5}
\end{figure}

For a deeper investigation of these dynamical indicators and to allow
a direct comparison with observations, we removed the explicit
dependence on time by plotting one parameter against another. The
three possible combinations are shown in Figures \ref{fig:a5p5},
\ref{fig:a5s2.5}, and \ref{fig:p5s2.5}, where the dashed lines are the
polynomial fits to the distributions of points in the BF0 case (left
panels), and they are reported for reference also for the BF10 and
BF20 runs in the central and right panels, respectively.  As shown by these 
figures, the measured values gradually move from the bottom-left to
the top-right corner of each diagram for increasing dynamical age, up
to CC (i.e., from green, to cyan, to blue colors). Then, in the
post-CC stage (yellow symbols) they tend to be smaller than or mixed
with those obtained during CC.  The point distributions follow
essentially the same relation (dashed lines in the figures) in all the
simulations, irrespective of the binary fraction.  In principle, then,
just from the measure of two parameters, these diagrams allow one to
understand whether a stellar system is in an early, intermediate or
advanced stage of dynamical evolution, although the range of values
sampled by the parameters decreases for increasing binary fraction,
due to the milder contraction of the core.  This illustrates the
complexity of univocally deriving the internal dynamical stage of the
cluster if the primordial binary fraction is unknown.  Nevertheless,
these diagrams also show that some useful information can still be
obtained. In fact, irrespective of the binary fraction, the
bottom-left corner of these plots is exclusively populated by
dynamically young systems (green symbols), and the largest values
found for the BF20 run correspond to quite evolved systems also in the
other cases. In addition, if in an observed cluster one measures
values of $A_5$, $P_5$ and $S_{2.5}$ that fall in the top-right corner
of these diagrams, a large binary content can be excluded, and the cluster 
is likely to be in a quite advanced stage of dynamical evolution.


\section{Zooming on the deviations from King models}
\label{sec:cfr_king}
\begin{figure}
\centering \includegraphics[scale=0.4]{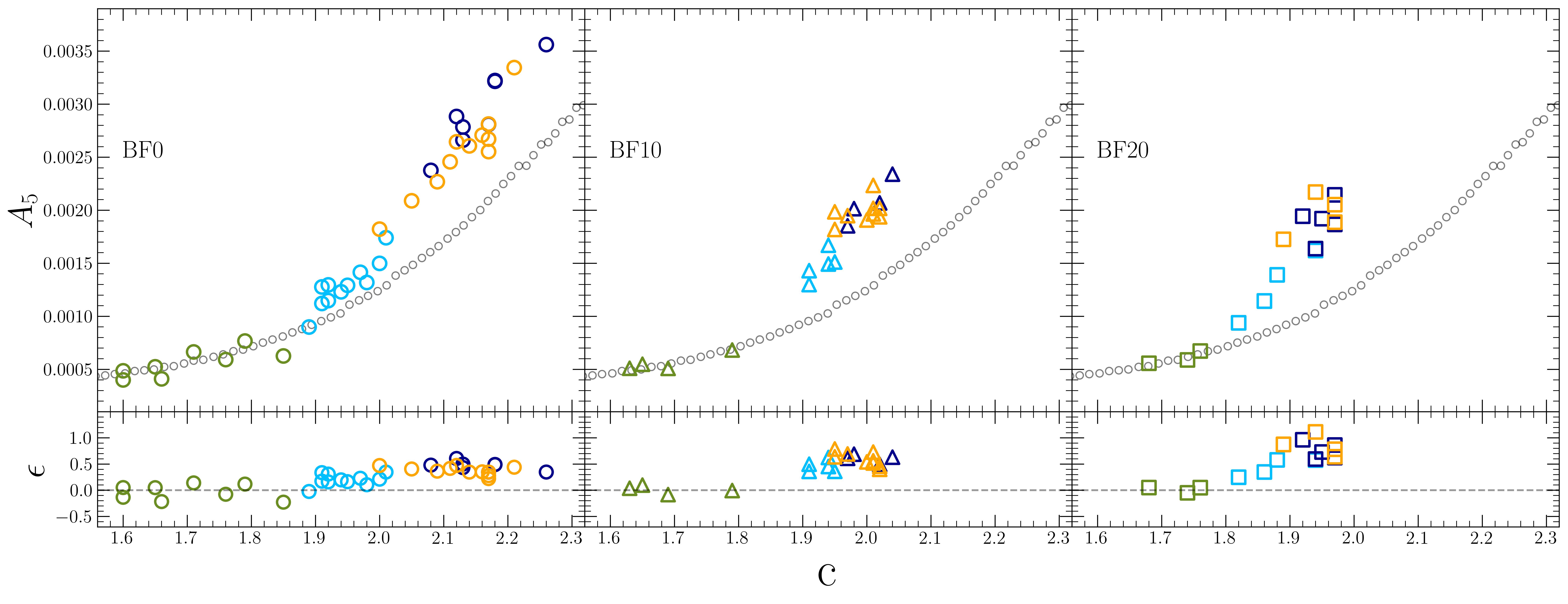}
\caption{Values of $A_5$ measured from the direct integration of the
  King model density profile for different values of the concentration
  parameter $c$ (black empty circles), compared to those obtained from
  the nCRDs of the extracted snapshots (colored symbols, the same as
   in the top row of in Fig. \ref{fig:all_params}) plotted
  against the concentration parameter of the King models that best-fit
  the density profile of each simulation.  The left, central and right
  panels refer to the BF0, BF10, and BF20 simulations, respectively.
  The bottom panels show the relative difference between the two
  measurements: $\epsilon=(A_5-A_5^{\rm King})/A_5^{\rm King}$ }
\label{fig:a5c}
\end{figure}

By definition, the nCRD traces the projected radial distribution of
stars with respect to the cluster center. Thus, it strictly depends on
the projected density distribution of the system. In turn, the latter
is commonly described through the King model family, which provides
very good fits for clusters in early dynamical stages, while show
significant discrepancies in the central regions for dynamically
evolved systems (see Section \ref{sec:dens} and, e.g., Figure 4 in
B22). We therefore built the nCRDs by directly integrating the King
model density profile for different values of the concentration
parameter $c$ between $\sim 1$ and 2.5, and we used these functions to
measure the three dynamical indicators defined above. The results
obtained for the $A_5$ parameter are shown in Figure \ref{fig:a5c} as
empty black circles. The colored symbols in the figure correspond to
the values determined from the actual nCRDs of the simulations, built
as described in Section \ref{sec:A5etc}, plotted as a function of the
concentration parameter of the King models that best-fit the
``observed'' density profiles for each extracted snapshot.  The
bottom panels show the relative difference between the parameter
obtained from the actual nCRD and those obtained from the King model
integration: $\epsilon=(A_5-A_5^{\rm King})/A_5^{\rm King}$.  This
figure shows that, as expected from the gradual increase in
concentration of the system, the values of both $A_5$ and $c$ increase
from early (green colors), to intermediate (cyan), to evolved
dynamical stages (blue and yellow).  In addition, in the very early
phases of dynamical evolution (green symbols), the density
distribution is properly reproduced by the King family, and the $A_5$
parameter measured from the actual nCRDs is essentially the same as
that obtained from the King models, irrespective of the primordial
binary fraction.  Then, from the pre-CC stage (cyan) onwards, the two
measures of $A_5$ start to differ: in fact, the growth of the stellar
density toward the center of the system makes the nCRDs increasingly
steeper, and the value of $A_5$ starts to systematically and growingly
exceed the corresponding value of $A_5^{\rm King}$.  Hence, the
  $A_5$ parameter can be used as a sort of magnifier to pinpoint
  clusters with density profile deviating from the King model
  distribution, well before the contraction of the system produces a
  measurable central cusp in the observed profile.

\section{Effect of dark remnants on the nCRD dynamical indicators}
\label{sec:dark_remnants}
\begin{figure}
\centering \includegraphics[scale=0.4]{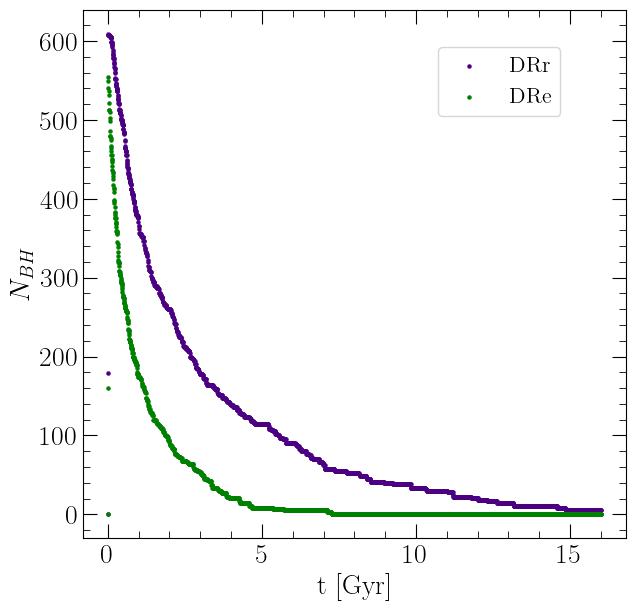}
\caption{Time evolution of number of black holes in the two simulations
  with a large retention fraction of dark remnants: DRr (indigo circles) and DRe (green circles).}
\label{fig:time_nbh_dark}
\end{figure}

While the previous sections focused on the effects of primordial binary systems,
here we discuss how the nCRD dynamical indicators are affected by the
presence of dark remnants. To this end, we consider the DRr and DRe simulations where a reduced kick velocity for black
holes is adopted according to the fallback prescription of
\citet{belczynski+02}, thus significantly enhancing the fraction of BHs retained within the system potential well, with respect to the BF0, BF10, and BF20 simulations. The time evolution of the number of black holes for the DRr and DRe runs is shown in Figure \ref{fig:time_nbh_dark}. As apparent, the rate of ejection of black holes in the DRe simulation (green circles) is much higher than that of the DRr simulation (indigo circles). 

\begin{figure}
\centering \includegraphics[scale=0.4]{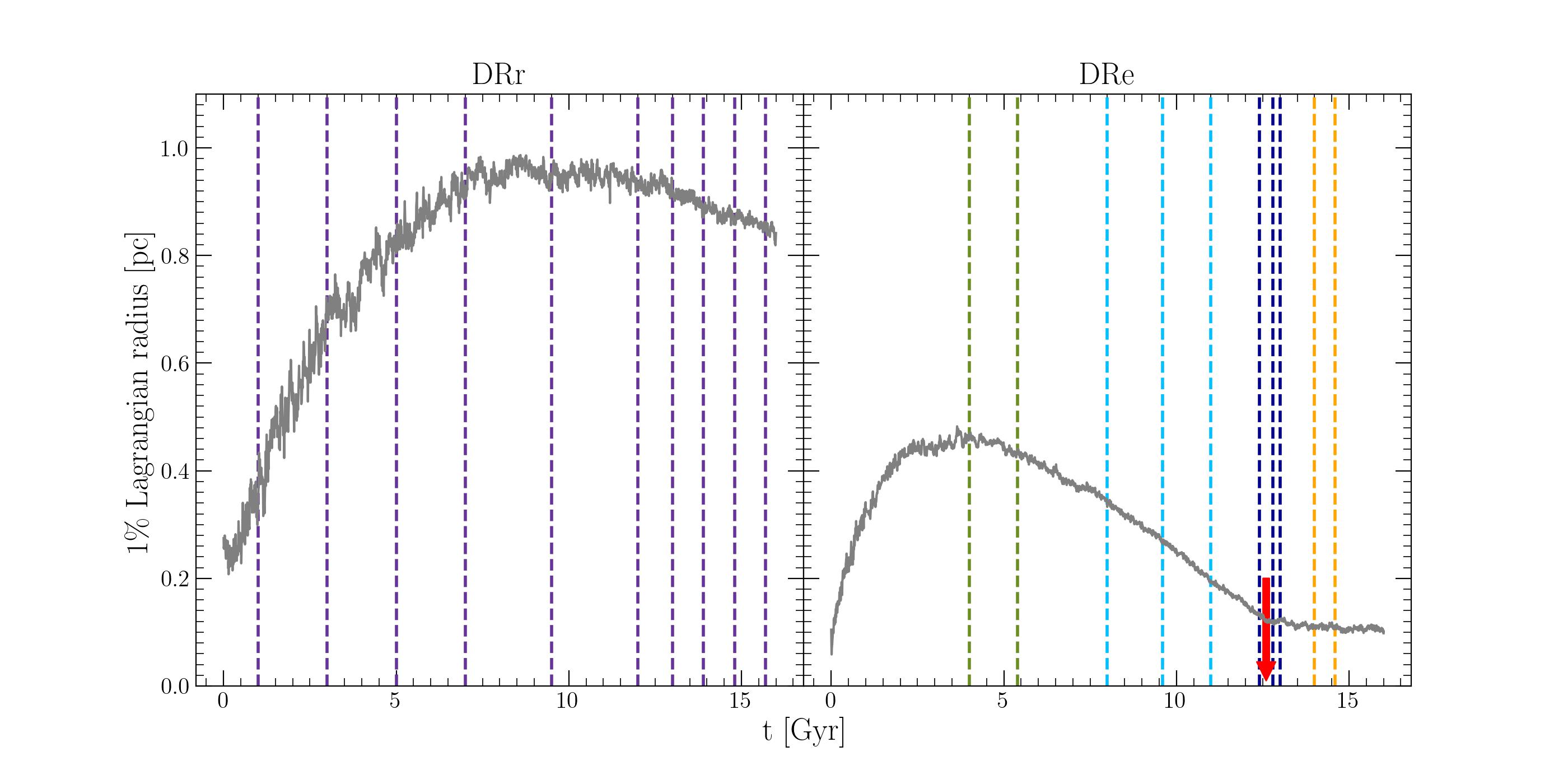}
\caption{Time evolution of 1\% Lagrangian radii in the two simulations
  with large retention fraction of dark remnants, DRr and DRe runs on	left and right panels respectively. The vertical dashed lines in indigo color mark the ten time snapshots extracted from the DRr simulation and the vertical dashed lines with standard color code used in previous figures are the ten time snapshots extracted from DRe simulation.}
\label{fig:lagrange_dark}
\end{figure}

 Figure \ref{fig:lagrange_dark} shows the time evolution of 1\%
Lagrangian radius of the DRr and DRe runs (on left and right panel respectively), with the extracted time snapshots
marked by vertical dashed lines. At odds with the simulations
discussed so far (compare to Fig. \ref{fig:lagrange_rad}), the
evolution of $r_{1\%}$ in the DRr simulation shows an extended expansion phase lasting up to about 10 Gyr driven by the retained black holes.

Such a long expansion
phase is then followed by a gradual decrease of $r_{1\%}$, where
two-body relaxation drives the cluster contraction. The system, 
however, never reaches the most advanced dynamical
phases and the CC stage in a Hubble time. In the case of the DRe simulation (right panel of Fig \ref{fig:time_nbh_dark}), due to its higher rate of ejection, most of the black holes are ejected by $\sim$ 5 Gyr and, after that time, the cluster's inner regions  contract until the collapse is halted by primordial binaries at about 13.7 Gyr;  the post-CC stages are similar to the BF10 and BF20 models.

\begin{figure}
\centering
\includegraphics[scale=0.4]{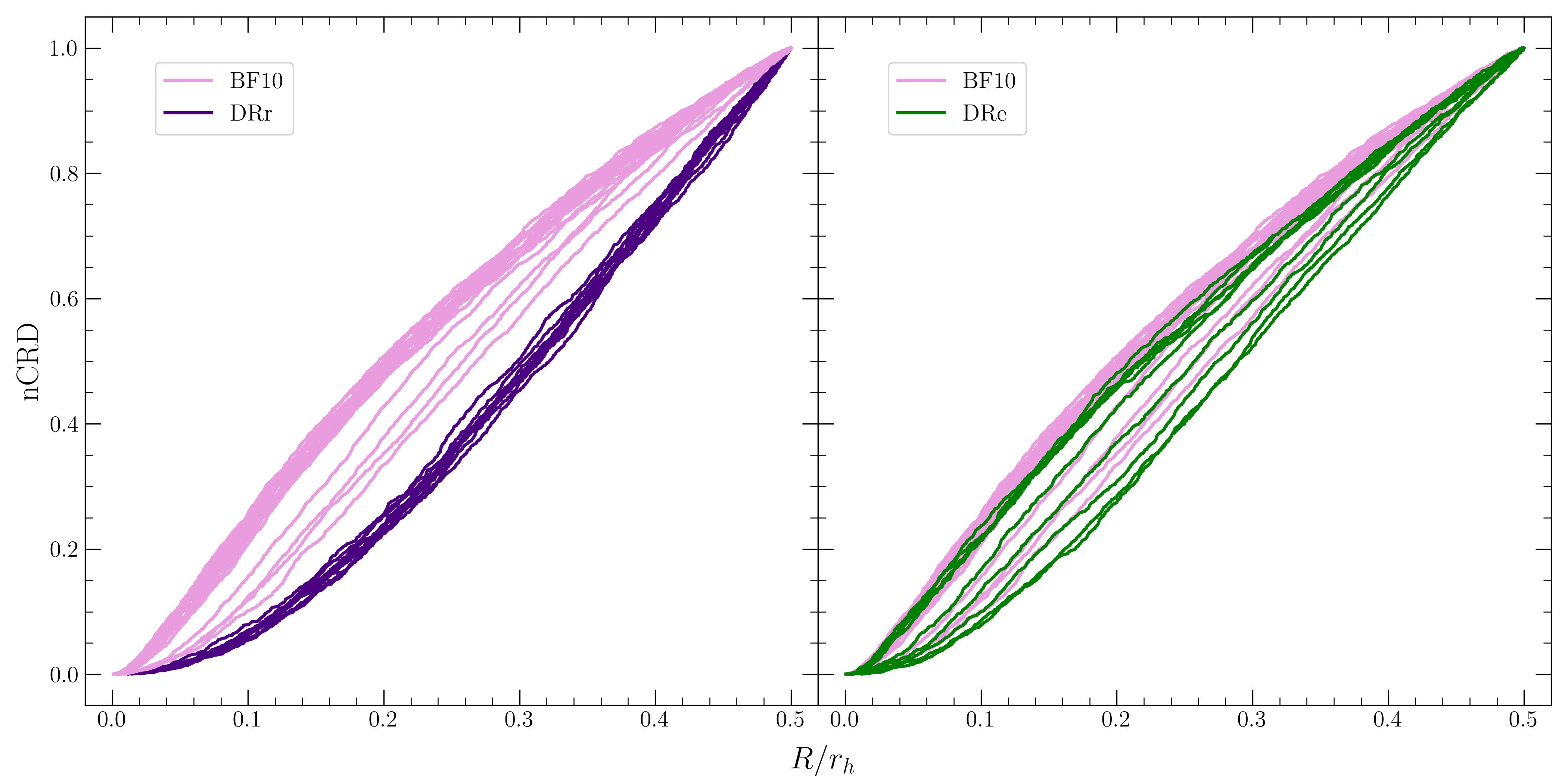}
\caption{Normalised cumulative radial distributions for all the
  snapshots extracted from the DRr (indigo lines) and DRe (green lines) simulations on the left and the right panels respectively. They are compared
  to those obtained for the BF10 run (pink lines, the same as in
  the right panel of Fig. \ref{fig:crds}).}
\label{fig:crd_dark}
\end{figure}

\begin{figure}
\centering
\includegraphics[scale=0.4]{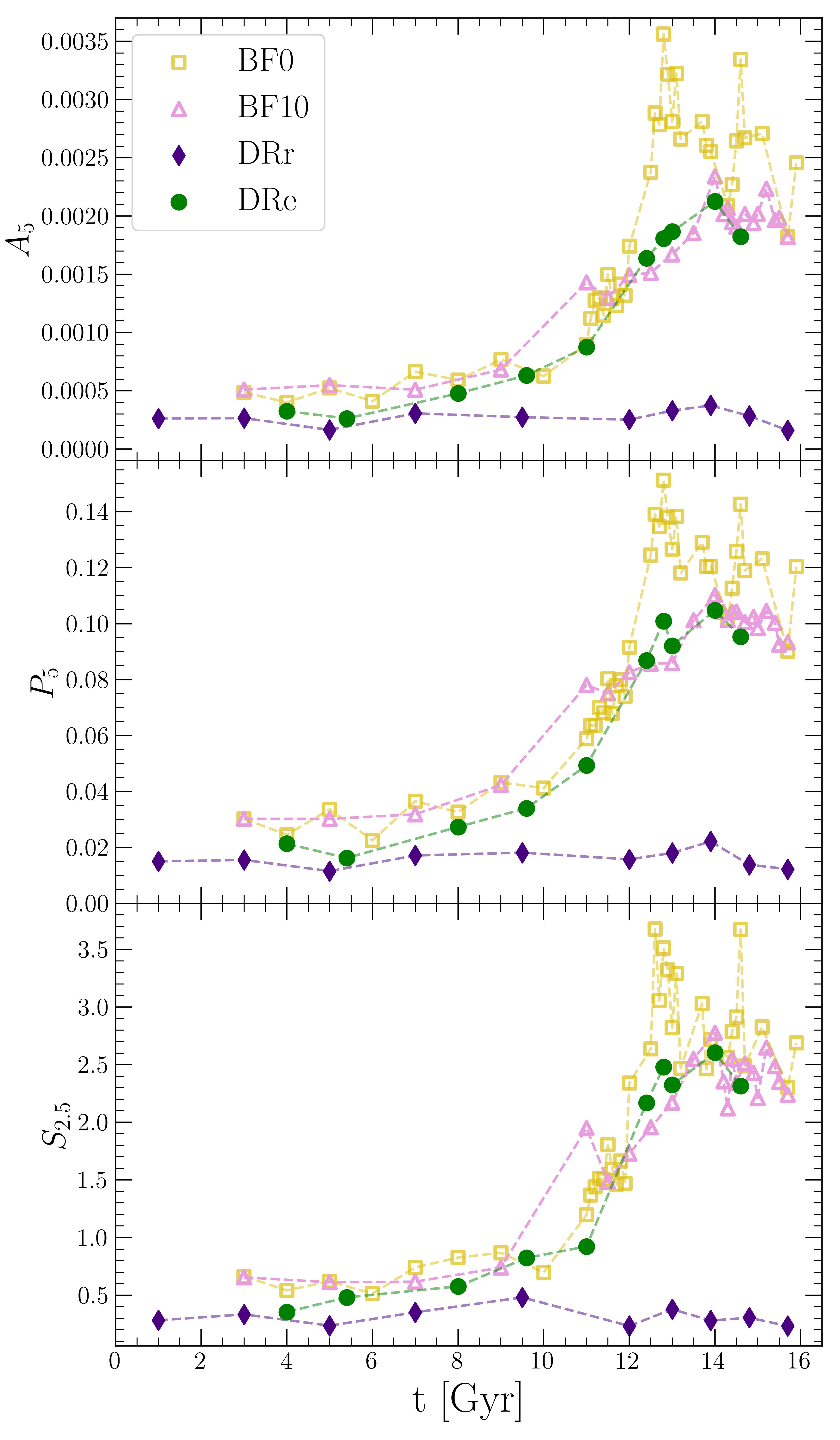}
\caption{Time evolution of the $A_5$, $P_5$ and $S_{2.5}$ parameters
  for the DRr run (indigo diamonds) and the  DRe run (green circles), compared to those measured in the
  BF10 and BF0 simulations (pink triangles and gold squares,
  respectively).}
\label{fig:params_dark}
\end{figure}

By adopting the same procedures described in Section \ref{sec:crd}, we
computed the nCRDs for all the extracted snapshots  of DRr and DRe. The nCRDs for DRr are shown in indigo lines and those of DRe in green lines on the left and right panels of Figure \ref{fig:crd_dark} respectively. In the same figure, they are compared to the nCRDs obtained from the BF10 run (pink lines). At all evolutionary times,  the nCRDs corresponding to the DRr run are all clumped (hence, they essentially
show the same morphology) and have much shallower slopes than the  BF10
run, implying a much smaller percentage of stars also in the innermost
regions, where the three dynamical indicators are defined. This is
fully confirmed by the time dependence of $A_5$, $P_5$, and $S_{2.5}$
shown in Figure \ref{fig:params_dark}, where the values obtained from
the DRr run (indigo diamonds) are compared to those measured in the  BF10
and BF0 simulations (pink triangles and gold squares,
respectively): the former always show much smaller values than the
others, and they are almost independent of time, consistently with the
lack of a significant structural evolution displayed by $r_{1\%}$ in
Fig. \ref{fig:lagrange_dark}. The shapes of the nCRDs of DRe evolve from having shallower slopes in the early snapshots (which are still affected by the expansion induced by the retained BHs), to having shapes indistinguishable from those of BF10 (at $t>= 7$ Gyr, when essentially all the BHs have been ejected and the evolution becomes similar to that of BF10). Indeed, the time dependence of the dynamical indicators shown in Figure \ref{fig:params_dark} for the DRe run (green circles) clearly starts from a low value similar to DRr run and follows an increasing trend like in BF0 and BF10 runs. In particular they are closer to the values of BF10 run. This reflects the fact that even though the cluster starts with a significant number of dark remnants (hence the nCRDs are shallow and the dynamical indicators have low values, like in DRr run), as the cluster looses all the black holes it undergoes standard dynamical evolution and the three dynamical indicators follow the trend of BF10 simulation which has no dark remnants. 

\section{Discussion and summary}
\label{sec:discussion}
In this paper we have investigated the effects of different primordial
binary fractions and dark remnant content on the value of three
parameters specifically defined by B22 to quantify the dynamical
evolution of the structural properties of star clusters.  The
(expected) impact of the fraction of binaries and dark remnants on the
dynamical evolution of the system translates in a corresponding effect
on the time dependence of the three parameters.

In the  DRr (where the rate of BH ejection is very low)  run, the expansion effect due to the retained black hole
system dominates most of the cluster's dynamical evolution.  As a
consequence, no significant contraction of the cluster occurs (see
 left panel of Fig. \ref{fig:lagrange_dark}), and the values of the three parameters
remain essentially constant with time and always much smaller than
those measured in  all the other simulations (see
Fig. \ref{fig:params_dark}), where the number of BHs is below a few units either from the beginning (BF0, BF10, and BF20 runs), or for most of the cluster life (DRe simulation). Hence, although in this case the nCRD
parameters cannot help understanding the dynamical age of the cluster
 (also because it does not go across well-distinct dynamical
phases), measuring such low values might be used as an indication that
the system  still includes a significant number of stellar-mass black
holes (or, at least, that it retained a significant number of dark remnants for a large fraction of its life).  The DRe run, which started with the same number of BHs as in the DRr simulation, but then ejected almost all of them during the initial $\sim 5$ Gyr of its evolution, essentially behaves as the models with no dark remnant retention. This implies that, while the three nCRD parameters cannot help disentangling among DRe and these latter cases, they allow us to determine the dynamical stage of star clusters with a high rate of BH ejection, even if they initially retained a large amount of these compact objects.
 Of course, however, deeper investigations are needed, and this
will be addressed in a forthcoming paper, where we will also
investigate the impact of a central intermediate-mass black hole.

In the three models with different primordial binary fractions and  no BH retention, the
general trend is that the values of the three parameters gradually
increase with time, and trace the internal dynamical evolution of the
cluster. The time variations of $A_5$, $P_5$, and $S_{2.5}$ are most
pronounced in the BF0 simulation, then become less marked in systems
including primordial binary stars reflecting the milder contraction of
these clusters.  Strictly speaking, this implies that to properly
infer the dynamical age of a cluster, one needs not only to measure
the values of these parameters, but also know the binary fraction of
the system.  Nevertheless, some general conclusions could be drawn
even if this information is not available. In fact,
 Figs. \ref{fig:all_params}$-$\ref{fig:p5s2.5} show that low values of
the three parameters ($A_5\lsim 0.0006$, $P_5\lsim 0.04$, and
$S_{2.5}\lsim 0.08$) indicate young dynamical ages, irrespective of
the binary content of the system.  In addition, large values of the
three parameters (as $A_5\gsim 0.0023$, $P_5\gsim 0.11$, and
$S_{2.5}\gsim 2.7$) would univocally indicate a cluster in an advanced
stage of dynamical evolution with no or very little primordial
binaries.  Indeed, the top-right corner of these diagrams is always
empty in the BF10 and BF20 simulations, and this can be used as an
indirect evidence against a large binary fraction in the system under
analysis.  Also the comparison between the values
of $A_5$ measured from the nCRD and those expected from the King model
distribution (Fig. \ref{fig:a5c}) shows that this parameter is able to
trace the progressive deviations from the King model expectations that
occur during the late stages of dynamical evolution.   Hence, in both theoretical and observational studies, A5 can be used as a sort of magnifier to
  identify a dynamically old system well before its contraction
  produces a measurable central cusp in the density profile.

The analysis presented in this paper confirms that the three nCRD
parameters introduced in B22 are very useful tools to investigate the dynamical stage of
stellar systems, even in the cases of a non-zero primordial binary
fraction  and in the case of clusters with a high rate of BH ejection. In this investigation the analyzed simulations follow the
evolution of a given system over the cosmic time, while the GCs in our
galaxy all have essentially the same, old age ($\sim 12$ Gyr), but are
observed in different stages of their internal dynamical evolution. A
forthcoming paper (Bhavana Bhat et al. 2022, in preparation) will be
thus devoted to the analysis of the nCRD parameters in a sample of
simulated clusters generated from a broad range of different initial
conditions (hence, subject to different rates of internal dynamical
evolution), but all considered at the same chronological age of $\sim
12$ Gyr.  Meanwhile, we are also performing detailed measures of the
nCRD parameters in real clusters, also discussing the sensitivity of
these new tools with respect to other dynamical indicators, in
particular, the $A^+$ parameter measured from the blue straggler star
population \citep[see][]{alessandrini+16, ferraro+12, ferraro+18,
  ferraro+19, ferraro+20}. This will finally provide a quantitative
assessment of the operational ability of the three nCRD diagnostics to
distinguish star systems in different stages of dynamical evolution.

\section*{acknowledgments}  
This work is part of the project Cosmic-Lab at the Physics and
Astronomy Department "A. Righi" of the Bologna University
(http://www.cosmic-lab.eu/ Cosmic-Lab/Home.html). The research was
funded by the MIUR throughout the PRIN-2017 grant awarded to the
project Light-on-Dark (PI:Ferraro) through contract PRIN-2017K7REXT.

\bibliography{arxiv_paper2}{}
\bibliographystyle{aasjournal}

\end{document}